\documentclass[showpacs,10pt,twocolumn,prx]{revtex4-1}
\usepackage{amsmath}
\usepackage{amssymb}
\usepackage{graphics}
\usepackage{epsfig}
\usepackage{CJK}
\usepackage{color}
\usepackage{lineno}

\begin{document}

%\begin{CJK*}{GBK}{}

\title{Extreme anisotropy and anomalous transport properties of heavily electron doped Li$_{x}$(NH$_{3}$)$_{y}$Fe$_{2}$Se$_{2}$ single crystals}
\author{Shanshan Sun$^{1,\dag}$, Shaohua Wang$^{1,\dag}$, Rong Yu$^{1}$, and Hechang Lei$^{1,*}$}
\affiliation{$^{1}$Department of Physics and Beijing Key Laboratory of Opto-electronic Functional Materials $\&$ Micro-nano Devices, Renmin University of China, Beijing 100872, China}
\date{\today}

\begin{abstract}
The missing hole packets near the Brillouin zone center render unique electronic structures to the heavily electron doped FeSe-based superconductors with $T_{c}$ above 40 K. It challenges the existing scenario accounting for the nature of superconductivity in the iron-based family. Yet, one hurdle that has to be overcome is the materials complexity in the rather limited number of compounds. Here we report the growth of heavily electron doped Li-NH$_{3}$ intercalated FeSe single crystals that are free of such complexities and allow access to the intrinsic superconducting properties. Our results show extremely large electronic anisotropy in both normal and superconducting states. Moreover, the anomalous transport properties appear in normal state, which are believed related to the anisotropy of relaxation time and/or temperature dependent electron carrier concentration. In view of the great chemical flexibility of intercalants, our findings provide novel platform to understanding of superconductivity origin of FeSe-related superconductors.
\end{abstract}

\pacs{74.70.Xa, 74.25.Sv, 74.25.Op, 74.25.-q}
\maketitle

\section{Introduction}

The electronic structure of FeAs-based superconductors (SCs) usually consists of hole pockets near the Brillouin zone center ($\Gamma$ point) and electron pockets near the Brillouin zone corners ($M$ point) \cite{Stewart}. The electron scattering between the electron and hole pockets gives rise to a sign change $s$-wave pairing (denoted by $s^{\pm}$ pairing) \cite{Mazin,Kuroki}, as proposed by the spin-fluctuation-mediated pairing theory. Such an $s^{\pm}$ pairing is consistent with most experimental observations \cite{HongDing,PCDai_RMP}. As for the FeSe-based SCs, experimental evidences suggest a quite similar pairing mechanism for the bulk FeSe and Fe(Te, Se) \cite{LiuX}. However, the discovery of heavily electron doped (HED) FeSe-based SCs, including A$_{x}$Fe$_{2-y}$Se$_{2}$ (A = alkali metals and Tl), monolayer FeSe thin films grown on SrTiO$_{3}$ (m-FeSe/STO films), and the recently discovered (Li, Fe)OHFeSe (FeSe-11111), raises a great challenge to the aforementioned pairing mechanism. On the one hand, from transport and spectral measurements \cite{Zhao2,ZhangWH,GuoJG,Lu}, these HED FeSe-based SCs show a significant enhancement of $T_{c}$ to 30 - 40 K compared to bulk FeSe at ambient pressure ($T_{c}\approx$ 8 K)\cite{Hsu}; One the other hand, the hole pockets, which are crucial in the above pairing mechanism, are absent in these compounds \cite{Zhao2}. Thus, the mechanism of high $T_{c}$ superconductivity in these HED FeSe-based SCs is still unsettled. However, the limited number of reported compounds with various material complexities make a systematic study on their intrinsic properties difficult. For example, in A$_{x}$Fe$_{2-y}$Se$_{2}$, the superconductivity could be affected by the iron vacancy ordering in the concomitant insulating phase with chemical formula A$_{0.8}$Fe$_{1.6}$Se$_{2}$ \cite{BaoW,ChenF}. In m-FeSe/STO films, the enhancement of $T_{c}$ from the bulk FeSe could be associated with the interface \cite{WangQY,LeeJ,Zheng}. With regard to FeSe-11111, the canted antiferromagnetic order in (Li, Fe)OH layers could have some influences on the superconductivity in FeSe layers \cite{Lu}. Therefore, in order to confirm that high $T_c$ is a universal property of the HED FeSe-based SCs, it is highly desired to find a compound without any material complexity.

In recent, superconductivity with $T_{c}$ up to 45 K has been reported in AM$_{x}$(NH$_{3}$)$_{y}$Fe$_{2}$Se$_{2}$ (AM = alkali, alkali-earth, and rare-earth metals) \cite{Ying,Scheidt,Ying2,Lucas,Sedlnaier,Lei1}. In these materials, the Fe vacancies are almost absent in FeSe layers \cite{Ying2,Lei1}, and the AM-NH$_{3}$ layers does not exhibit magnetic order. Thus, they serve as a good model system to study the superconductivity of HED FeSe-based SCs. But the investigation of their intrinsic physical properties, especially transport properties, is impeded because of the limited size of (00l)-orientated parent FeSe single crystals, the difficulty of completed intercalation of FeSe single crystals at low temperature as well as the unstability of samples at room temperature.

In this work, we grown Li$_{x}$(NH$_{3}$)$_{y}$Fe$_{2}$Se$_{2}$ (LiFeSe-122) single crystals successfully and report a comprehensive study on their transport properties. We find that intercalation of Li-NH$_{3}$ significantly enlarges inter-FeSe-layer distance $d$, leading to the extreme anisotropy of LiFeSe-122 in both normal and superconducting states. Such large $d$ with weak interlayer interaction suggests that LiFeSe-122 would be more two-dimensional (2D) and become a good reference to m-FeSe/STO films. The dominant electron-type carriers with rather high concentration confirms its HED feature and suggests that a universal high $T_{c}$ superconductivity persists in a rather wide range of carrier doping concentration in these family of SCs. Moreover, possibly due to the anisotropic relaxation time and the change of carrier concentration with temperature, LiFeSe-122 shows anomalous transport properties at normal state, implying the exotic Fermi surface topology. Our work will be helpful for understanding the underlying pairing mechanism in the HED FeSe-based SCs. It will also shed light on the origin of superconductivity in other iron-based SCs.

\section{Experimental Method}

Single crystals of FeSe were grown by a chemical vapor transport technique using elemental Fe and Se and a eutectic mix of the chlorine salts. The Fe and Se powders were mixed together with the molar ratio of 1 : 1 and loaded into silica ampoules. Then the eutectic flux of AlCl$_{3}$ and NaCl (= 0.52 : 0.48) was mixed and loaded into the ampoules which were subsequently evacuated down to 1 Pa and sealed. The ampoules were put into the horizontal furnace and heated with temperature gradient (high temperature region: 690 K; low temperature region: 620 K) for 30 days. Finally, the furnace was shut down and the ampoules cooled down naturally. The tetragonal crystals can be obtained after dissolved in distilled water and rinsed in alcohol. The LiFeSe-122 single crystals were synthesized by the low-temperature ammonothermal technique. The FeSe single crystals and pieces of Li metal with nominal mole ratio of 1 : 2 as well as a magnetic stirrer were loaded and sealed into a designed high-pressure vessel (25 mL) with a stop valve for evacuating treatment and another one for introducing NH$_{3}$. These manipulations were carried out in an argon-filled glove box in order to prevent air and water contamination. Then, the vessel was taken out from the glovebox and evacuated down to 1 mPa by a turbo molecular pump. The NH$_{3}$ gas was introduced and condensed by cooling the vessel to 238 K for 20 minutes. After that, the vessel was taken out from the cooling bath and stirred for 2 days at room temperature in order to facilitate the reaction and to improve the homogeneity of intercalation. The crystals with typical size 2$\times$$3\times$0.05 mm$^{3}$ can be obtained after evacuating the NH$_{3}$ gas. X-ray diffraction (XRD) patterns were collected using a Bruker D8 X-ray Diffractometer with Cu $K_{\alpha}$ radiation ($\lambda=$ 0.15418 nm) at room temperature. The elemental analysis was performed using the inductively coupled plasma atomic emission spectroscopy (ICP-AES) and energy-dispersive X-ray spectroscopy (EDX) analyses. Electrical transport measurements were carried out in a Quantum Design PPMS-14. The longitudinal and Hall electrical resistivity were measured using a four-probe method on rectangular shaped single crystals. The current flows in the $ab$ plane of crystal. The Hall resistivity was obtained from the difference of the transverse resistivity measured at the positive and negative fields in order to remove the longitudinal resistivity contribution due to voltage probe misalignment, i.e., $\rho_{xy}(\mu_{0}H)=[\rho(+\mu_{0}H)-\rho(-\mu_{0}H)]/2$. The $c$-axis resistivity $\rho_{c}(T)$ was measured by attaching current and voltage wires on the opposite sides of the plate-like crystal. Magnetization measurements were performed in a Quantum Design Magnetic Property Measurement System MPMS3.

\section{Results}

\begin{figure}
\centerline{\includegraphics[scale=0.21]{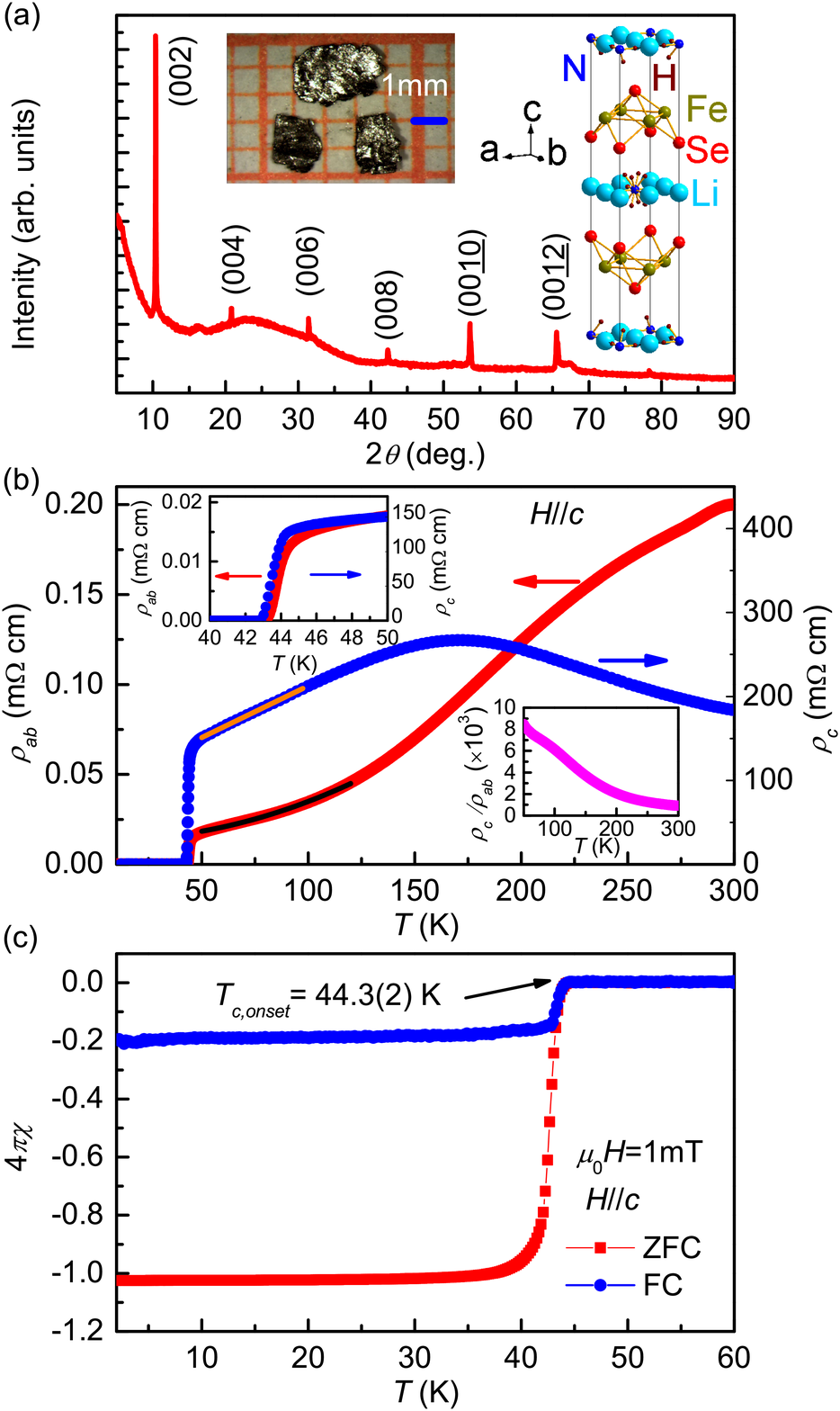}} \vspace*{-0.3cm}
\caption{Characterization of structure, zero-field transport and magnetization properties of LiFeSe-122 single crystals. (a) XRD pattern of a LiFeSe-122 single crystal. Insets: (a) photo (left) and (b) crystal structure of LiFeSe-122 single crystals. The length of one grid in the photo is 1 mm. (b) Temperature dependence of in-plane resistivity $\rho _{ab}(T)$ and $c$-axial resistivity $\rho _{c}(T)$ at zero field. Inset: (a) enlarged resistivity curve near $T_{c}$ (top left) and (b) the ratio of $\rho _{c}/\rho _{ab}$ as a function of temperature. (c) Temperature dependence of dc magnetic susceptibility $4\pi\chi$ up to 60 K in $\mu_{0}H=$ 1 mT with ZFC and FC modes.}
\end{figure}

%\textbf{Structure, resistivity anisotropy and the superconducting transition.}
The crystal structure of LiFeSe-122 consists of FeSe layers intercalated by the Li-NH$_{3}$ layers (inset (b) of Fig. 1(a)). It has the same space group as K$_{x}$Fe$_{2-y}$Se$_{2}$ at $T>T_{N}$ (I4/mmm) \cite{BaoW}. In each unit cell, there are two layers of FeSe. The difference is K ions in K$_{x}$Fe$_{2-y}$Se$_{2}$ are located in the $2a$ position which are occupied by N atoms in LiFeSe-122, and Li ions are located in $2b$ and $4c$ positions in the latter. In contrast, the space group of FeSe-11111 is P4/nmm and there is only one layer of FeSe in each unit cell \cite{Lu}. Moreover, the Li-NH3 layers should not have magnetic order when compared to (Li, Fe)OH layers in FeSe-11111. The XRD pattern of a single crystal (Fig. 1(a)) reveals that the crystal surface is parallel to the $(00l)$-plane and the diffraction peaks shift to lower angle compared to FeSe single crystal (see Supplemental Material \cite{SI}). The thin plate-like crystals with square shape (inset (a) of Fig. 1(a)) is consistent with the layered structure of LiFeSe-122. The refined $a$- and $c$-axial lattice constants by using the powder XRD pattern (see Supplemental Material \cite{SI}) is 3.7704(6) and 16.973(7) \AA, respectively, consistent with the reported results \cite{Lei1,Scheidt,Lucas}. The $a$-axial lattice constant is close to those in the bulk FeSe (3.76 \AA) \cite{Hsu}, suggesting that the intercalation of Li-NH$_{3}$ rarely changes the in-plane chemical environment of FeSe layers \cite{Lei1}. In contrast, the distance $d$ between two adjacent FeSe layers ($\sim$ 8.487 \AA, equals half of $c$-axial lattice constant) is much larger than that in bulk FeSe (5.48 \AA) \cite{Hsu}, and somewhat smaller than that in FeSe-11111 (9.16 \AA) \cite{Lu}. This implies a rather weak interaction between the two adjacent FeSe layers, similar to FeSe-11111. Such a weak inter-FeSe-layer bonding is also evidenced by the easy cleavage of crystals along the $ab$-plane. The atomic ratio of Li : Fe : Se determined from the ICP-AES analysis is 0.18 : 1 : 0.9 and the measurement of EDX by examination of multiple points on the crystals gives Fe : Se = 0.987 : 1, clearly indicating the absence of Fe vacancy in the LiFeSe-122 crystals. Based on the ratio of Li$_{0.36}$(NH$_{3}$)$_{y}$Fe$_{2}$Se$_{2}$, the estimated electron doping level is $\sim$ 0.18 e/Fe.

Fig. 1(b) shows the temperature dependence of the in-plane resistivity $\rho_{ab}(T)$ and the out-of-plane resistivity $\rho_{c}(T)$ at zero field from 10 K to 300 K. The $\rho_{ab}(T)$ exhibits a metallic behavior in the whole measured temperature range and the residual resistivity ratio (RRR), defined as $\rho_{ab}$(300 K)/$\rho_{ab}$(50 K), is about 11.2. Between 50 K and 120 K, the $\rho_{ab}(T)$ curve can be well fitted using the formula $\rho_{ab}(T)=\rho_{0}+AT^{\alpha}$ with $\rho_{0}=$ 13.2(1) $\mu\Omega$ cm, $A=$ 1.5(1)$\times10^{-3}$ $\mu\Omega$ cm K$^{-2}$ and $\alpha=$ 2.07(2). When further lowering the temperature, a sharp superconducting transition appears in the $\rho_{ab}(T)$ curve at zero field at $T_{c,\rm{onset}}=$ 44.56(4) K and $\Delta T_{c}=$ 1.1 K (inset (a) of Fig. 1(b)). The $T_{c}$ is significantly higher than that of FeSe ($\sim$ 9 K, see Supplemental Material \cite{SI}). In contrast, there is a hump in the $\rho_{c}(T)$ curve with the peak position located around 170 K, i.e., the $\rho_{c}(T)$ shows an insulator behavior for $T>$ 170 K and metallic behavior at $T<$ 170 K. It has to be mentioned that the peak position varies from sample to sample in between 120 K and 170 K (see Supplemental Material \cite{SI}). Between 50 K and 100 K, the temperature dependence of $\rho_{c}(T)$ is nearly linear, obviously different from that in $\rho_{ab}(T)$. The $T_{c,\rm{onset}}$ for $\rho_{c}(T)$ at zero field is 44.1(1) K with $\Delta T_{c}=$ 1.2 K, consistent with that of $\rho_{ab}(T)$ (inset (a) of Fig. 1(b)). The sharp superconducting transitions for both current directions and rather large RRR in $\rho_{ab}(T)$ indicate the high quality of single crystal. On the other hand, the absolute value of $\rho_{c}(T)$ is much larger than $\rho_{ab}(T)$ at normal state. The ratio of $\rho_{c}$/$\rho_{ab}$ (inset (b) of Fig. 1(b)) is about 900 at 300 K and gradually increases to about 8000 when temperature is down to 50 K, indicating a significant anisotropy of LiFeSe-122 crystals. Notably, this anisotropy is even larger than that in FeSe-11111, although the former has a slightly smaller $d$ value \cite{Dong}. It suggests that the inter-FeSe-layer interaction might be even weaker in LiFeSe-122 than FeSe-11111.

The zero-field-cooling (ZFC) dc magnetic susceptibility $4\pi\chi(T)$ of LiFeSe-122 single crystal at $\mu_{0}H$ = 1 mT along the $c$-axis (Fig. 1(c)) shows that the superconducting shielding emerges at about 44.3(2) K with rather sharp transition width, consistent with the resistivity results and previous magnetic susceptibility measurements ($T_{c}\sim$ 44 K) \cite{Ying,Scheidt,Lucas,Lei1}. After considering the demagnetization effect of crystal, the superconducting volume fraction (SVF) estimated from the ZFC data is very close to 100 \%, indicating that the superconductivity in LiFeSe-122 single crystal is bulk. For field-cooling (FC) $4\pi\chi(T)$ curve, the SVF is about 20 \% at 2 K, implying the rather strong flux pinning effect in the crystal.

\begin{figure}
\centerline{\includegraphics[scale=0.21]{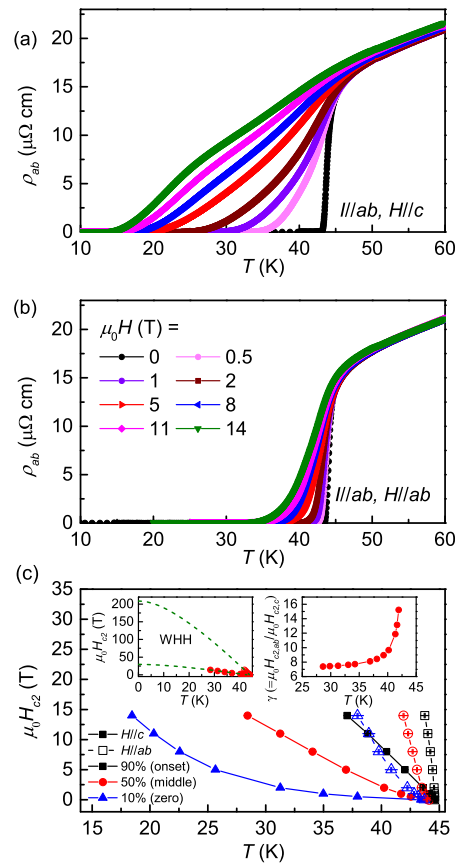}} \vspace*{-0.3cm}
\caption{Resistive upper critical fields of LiFeSe-122 single crystals. Temperature dependence of $\rho _{ab}(T)$ at various magnetic fields for (a) $H\Vert ab$ and (b) $H\Vert c$. (c) Temperature dependence of resistive upper critical fields $\mu_{0}H_{c2}(T)$ corresponding to three criterions for both field directions. Insets: (left) the WHH fitting of $\mu_{0}H_{c2}(T)$ (dashed lines) for both field directions and (right) the temperature dependence of the anisotropy of the upper critical field $\gamma=\mu_{0}H_{c2,ab}/\mu_{0}H_{c2,c}$ using the 50\% $\rho_{n}$ criterion.}
\end{figure}

%\textbf{Anisotropy of upper critical fields.}
Fig. 2(a) and (b) show the temperature dependence of $\rho _{ab}(T)$ at various magnetic fields up to 14 T for $H\Vert c$ and $H\Vert ab$, respectively. With increasing fields, the onset temperatures of superconducting transition only shift slightly to lower temperatures for both field directions. However, the transition width becomes broader at higher fields. The trend is more obvious for $H\Vert c$ than $H\Vert ab$. Such resistive tail for $H\Vert c$ has been observed in SmFeAsO$_{0.85}$ and cuprates \cite{Lee,Fendrich}, suggesting a wide vortex-liquid phase exists in LiFeSe-122 for $H\Vert c$. Interestingly, similar field-induced broadenings of resistive transitions have also been observed in FeSe-11111 \cite{Dong}. The upper critical fields $\mu_{0}H_{c2}(T)$ determined from the 90\%, 50\% and 10\% of $\rho_{n}$ in Fig. 2(a) and (b) are summarized in Fig. 2(c). The $\mu_{0}H_{c2}(T)$ curves for $H\Vert ab$ exhibit larger slopes than those for $H\Vert c$. Moreover, for $H\Vert c$ the $\mu_{0}H_{c2}(T)$ determined from 10\% $\rho_{n}$ is much smaller than those corresponding to 50\% and 90\% $\rho_{n}$ criteria because the region near 10\% $\rho_{n}$ is related to the vortex-liquid phase, while the region around 90\% $\rho_{n}$ is influenced by superconducting fluctuations. When choosing the criterion of 50\% $\rho_{n}$, the estimated $\mu_{0}H_{c2}(0)$ from the Werthamer-Helfand-Hohenberg (WHH) formula is 208.7 T and 29.5 T for $H\Vert ab$ and $H\Vert c$, respectively (inset (a) of Fig. 2(c)). Correspondingly, the anisotropy of $\mu_{0}H_{c2}$, $\gamma=\mu_{0}H_{c2,ab}/\mu_{0}H_{c2,c}$ is about 16 when $T_{c}$ is close to 44 K and then decreases with lowering temperature (inset (b) of Fig. 2(c)).

\begin{figure}
\centerline{\includegraphics[scale=0.23]{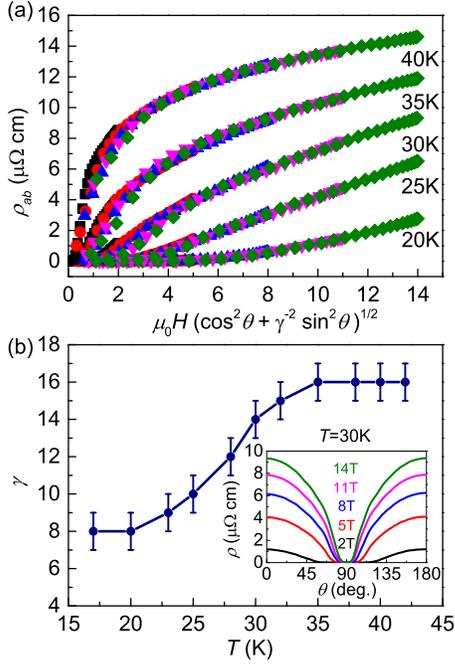}} \vspace*{-0.3cm}
\caption{Angular dependence of $\rho _{ab}(\theta, \mu _{0}H)$ at various fields for LiFeSe-122 single crystals. (a) Scaling behavior of $\rho _{ab}(\theta, \mu _{0}H)$ versus $\protect\mu_{0}H_{s}=\mu_{0}H(\cos^{2}\theta +\gamma ^{2}\sin^{2}\theta)^{1/2}$ from 20 K to 40 K at different magnetic fields. (b) The temperature-dependent anisotropy factor $\gamma$. Inset: angular dependence of $\rho _{ab}(\theta, \mu _{0}H)$ at 30 K with $\mu _{0}H$ = 2, 5, 8, 11, and 14 T.}
\end{figure}

In order to obtain a more accurate value of $\gamma$, the angular-resolved resistivity $\rho_{ab}(\theta, \mu _{0}H)$ under various fields at given temperatures have been measured (see Supplemental Material \cite{SI}). Taking the results at $T=$ 30 K as an example (inset of Fig. 3(b)), all of resistivity curves show a symmetric cup-like shape with the minimum value at $\theta=$ 90$^{\circ}$ and the maximum value at $\theta=$ 0$^{\circ}$ and 180$^{\circ}$, where $\theta$ is the angle between the direction of the external field and the $c$ axis. It indicates that the $\mu_{0}H_{c2,ab}$ is larger than the $\mu_{0}H_{c2,c}$. Moreover, the zero-resistivity region near $\theta=$ 90$^{\circ}$ becomes narrower with increasing field. According to the anisotropic Ginzburg-Landau theory $\mu_{0}H_{c2}^{GL}(\theta)=\mu_{0}H_{c2,ab}/(\sin^{2}\theta+\gamma^{2}\cos^{2}\theta)^{1/2}$ \cite{Blatter}, the $\rho_{ab}(\theta, \mu _{0}H)$ at a certain temperature under different magnetic fields can be scaled into one curve with a proper anisotropy parameter $\gamma=\mu_{0}H_{c2,ab}/\mu_{0}H_{c2,c}$. As shown in Fig. 3(a), by adjusting $\gamma(T)$, there are good scaling behaviors for LiFeSe-122 single crystals at various temperatures. The temperature dependence of $\gamma(T)$ (Fig. 3(b)) is similar to that shown in the right inset of Fig. 2(c), but the downward trend with decreasing temperature is milder than the latter. The values of $\gamma$ are much larger than those of bulk FeSe \cite{Vedeneev}, and even much larger than those in most of other iron-based SCs \cite{Hunte,Yuan HQ,Lei HC2}. However, they are comparable to those in FeSe-11111 \cite{Yi X}.

\begin{figure}
\centerline{\includegraphics[scale=0.17]{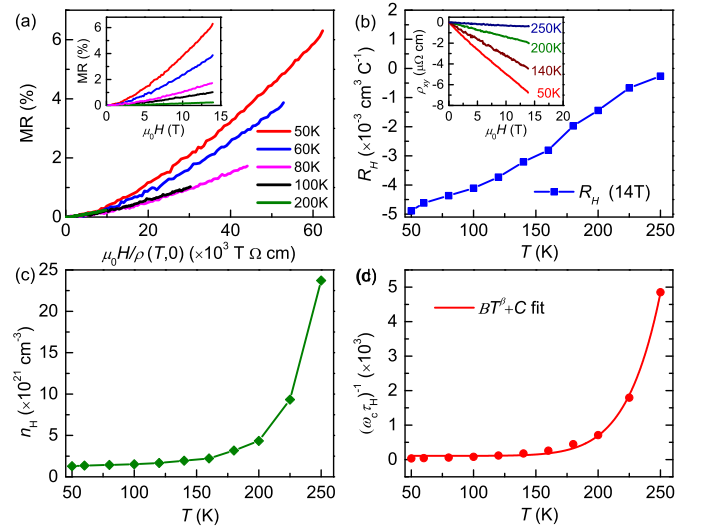}} \vspace*{-0.3cm}
\caption{Magnetoresistance and Hall effects of LiFeSe-122 single crystals. (a) Kohler plot between 50 - 200 K. Inset: field dependence of MR at different temperatures. (b) Temperature dependence of the Hall coefficient $R_{H}(T)$ at $\mu _{0}H$ = 14 T. Inset: field dependence of Hall resistivity $\rho_{xy}$ at various temperatures. (c) Temperature dependence of Hall number $n_{H}(T)=1/eR_{H}$. (d) cot$\theta_{H}$ as a function of temperature. The solid line is the fit by using the expression $BT^{\beta}+C$.}
\end{figure}

%\textbf{Anomalous transport properties at normal state.}
The transport properties at normal state reflects the electronic scattering and can tell information about Fermi surface (FS), which are important to understand the superconductivity in iron-based SCs. Inset of Fig. 4(a) shows the magnetoresistance (MR) [MR = [$\rho_{ab}(T,\mu _{0}H)$ - $\rho_{ab}(T,0)$]/$\rho_{ab}(T,0)$ = $\Delta\rho_{ab}$/$\rho_{ab}(T,0)$] at various temperatures. The MR is relatively small ($\sim$ 6.5 \% at 50 K and 14 T) and decreases gradually with increasing temperature. Usually, if there is an isotropic relaxation time $\tau$ at all points on the FS with a single type of carrier, the Kohler's rule will hold and the MR at different temperatures can be scaled by the expression MR = $f(\mu_{0}H\tau)=F(\mu_{0}H/\rho_{ab}{T,0})$ \cite{Pippard}. However, the Kohler's rule is clearly violated in LiFeSe-122 single crystal (Fig. 4(a)). In general, there are several possibilities leading to the invalidation of Kohler's rule, such as multiband effect, anisotropic $\tau$, and two different relaxation times related to zero-field resistivity and MR etc \cite{McKenzie}. In order to understand the origin of violation of Kohler's rule, the temperature dependence of Hall resistivity $\rho_{xy}$ are investigated. The $\rho_{xy}$ are in good linear relation against the magnetic field up to 14 T (inset of Fig. 4(b)) and the Hall coefficient $R_{H}=\rho_{xy}/\mu_{0}H$ at 14 T is negative in the whole range of measuring temperature (Fig. 4(b)). It indicates that the dominant carriers are electron-type, consistent with the electron doping of Li into FeSe layers. Moreover, $R_{H}$ becomes smaller with increasing temperature. Strikingly, the temperature dependence and even the absolute value of $R_{H}$ are quiet similar to those in gated-voltage-modulated FeSe and FeSe-11111 flakes with $T_{c}=$ 48 K and 43.4 K \cite{ChenXH1,ChenXH2}, strongly suggesting a universal underlying physics in these HED FeSe-based SCs. In contrast, FeSe-11111 exhibits the concave shape of $R_{H}$ and characteristic temperature $T^{*}\sim$ 120 K, possibly related to the magnetic properties of (Li, Fe)OH layers that are absent in LiFeSe-122 \cite{Dong}.

The angle-resolved photoemission (ARPES) results show that the hole pockets at $\Gamma$ point are absent in most of HED FeSe-based SCs, such as (Tl, Rb)$_{x}$Fe$_{2-y}$Se$_{2}$, FeSe-11111 and m-FeSe/STO films \cite{Zhao2}. Theoretical calculation on LiFeSe-122 also suggests that with electron doping the size of the hole pockets remarkably shrink while the size of electron pockets around the $M$ point increases \cite{Guterding}. Thus, the electron pockets in LiFeSe-122 should play a dominant role in the transport properties at normal state. As shown in Fig. 4(c), the apparent carrier concentration (i.e., the Hall number $n_{H}=1/eR_{H}$) is weakly temperature dependent below about 180 K, consistent with a single band model with only electron pockets in LiFeSe-122. The estimated $n_{H}$ at 50 K is about 1.3$\times10^{21}$ cm$^{-3}$, much larger than that in FeSe \cite{Watson}, but comparable $3\times10^{21}$ cm$^{-3}$, a value by assuming 0.18 electron is transferred from Li to each Fe atom. When $T>$ 180 K, the $n_{H}$ increases quickly, implying that other factors may have some influences on $n_{H}$, such as the anisotropic relaxation time $\tau$ at different points of FSs with different temperature dependence \cite{McKenzie,Hurd}, because the shapes of electron-type FSs for LiFeSe-122 are far away from cylindrical \cite{Guterding}. Moreover, the $n_{H}$ could also change at high temperature due to the thermal excitations of carriers among the narrow bands when Fermi energy level $E_{F}$ is close to the bottom or top of bands \cite{Brouet}.

The Hall angle cot$\theta_{H}= \rho_{xx}/\rho_{xy}\equiv 1/(\omega_{c}\tau_{H})$, where $\omega_{c}$ is the cyclotron frequency and $\tau_{H}$ is the relaxation time determined from Hall angle, can give more intrinsic feature of $\tau$ while separating the information about the carrier concentration away. For iron-based SCs, the cot$\theta_{H}$ usually follows the phenomenological function form cot$\theta_{H}=BT^{\beta}+C$ with $\beta$ values between 2 and 4 \cite{WangAF}. In contrast, the cot$\theta_{H}$ of LiFeSe-122 exhibits strong temperature dependence and increases quickly when $T>$ 180 K (Fig. 4(d)). The fit between 50 K and 250 K using above function gives $\beta =$ 9.3(3), which is enormously larger than other iron-based SCs. The origin of such large $\beta$ is still unclear and one of possible explanations for the significant enhancement of cot$\theta_{H}$ above 180 K could be the emergence of multiband effect because of thermal excitations of hole-type carriers at high temperature. Thus, both anisotropic $\tau$ accompanying with the possible multiband behavior at high temperature may lead to the violation of Kohler's rule in LiFeSe-122.

\section{Discussion}

\begin{figure}
\centerline{\includegraphics[scale=0.17]{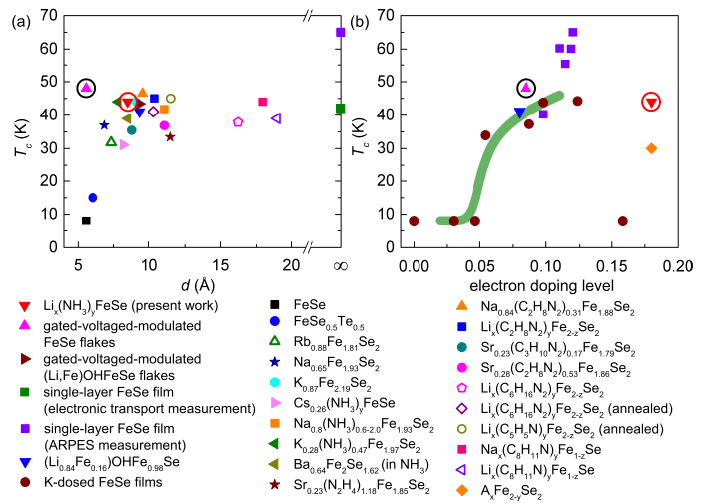}} \vspace*{-0.3cm}
\caption{Phase diagrams of HED FeSe-based SCs. $T_{c}$ as a function of (a) $d$ and (b) electron doping level in HED FeSe-based SCs. The results of present work and gated-voltaged-modulated FeSe flakes are emphasized by red and black circles, respectively. Data is taken from literature\cite{Zhao2,ChenXH1,ChenXH2,HeJ,Hatakeda,DLFeng}. The green line in (b) serves as guide to the eye.}
\end{figure}

The space group of both LiFeSe-122 and K$_x$Fe$_{2-y}$Se$_2$ is I4/mmm, while it is P4/nmm for FeSe-11111. It has been argued that this difference results in distinct electronic structures in the two classes of compounds, and will strongly affect the pairing symmetry of K$_x$Fe$_{2-y}$Se$_2$ \cite{Mazin2}. However, due to material complexity, the pairing symmetry of K$_x$Fe$_{2-y}$Se$_2$ has not been fully settled, though there are already a number of theoretical proposals \cite{Chubukov,JPHu,Nica}. Given the same structural symmetry and similar electron doping level, we expect the study on the superconducting pairing of LiFeSe-122 can solve this open issue. Despite the different lattice structures, both LiFeSe-122 and FeSe-11111 exhibit large anisotropy at both normal and superconducting states, and they have comparable $T_{c}$ \cite{Dong}. The electronic anisotropy is likely associated with the very large structural anisotropy (large $d$ values) among the iron-based SCs, which gives rise to a highly 2D electronic structure.

To understand the comparable $T_c$ in LiFeSe-122 to other HED FeSe-based SCs, we summarize the $T_c$ with the inter-FeSe-layer distance $d$ and the electron doping level in Fig. 5. %would suggest that such a quasi-2D electronic structure is crucial to the superconductivity in these compounds.}
Previous study summarizing the relationship between $T_{c}$ and $d$ value implies that $T_c$ will increase with increasing $d$ till $d\approx8.5$ \AA, above which $T_c$ gets to saturate \cite{Noji}. This seems to be valid for most of HED FeSe-based SCs (Fig. 5 (a)), but recent work on gated-voltage-modulated FeSe flakes with $T_{c}^{\rm{onset}}\sim$ 48 K (emphasized in Fig. 5(a) by black circle) \cite{ChenXH1} indicates that the $d$ value may not be the major factor to the enhancement of superconductivity in FeSe-based SCs. More importantly, notice that all high $T_{c}$ FeSe-based SCs at ambient pressure with increased $d$ values are HED.
%For instance, the carrier doping level is 0.18 e/Fe in LiFeSe-122 with $T_{c}\sim$ 44 K, 0.08 e/Fe in FeSe-11111 with $T_{c}\sim$ 41 K (ref. \mycite{Zhao2}), 0.12 e/Fe in vacuum-annealed m-FeSe/STO films with $T_{c}\sim$ 65 K determined from ARPES measurement\cite{He}, and 0.085 e/Fe in gated-voltage-modulated FeSe flakes with $T_{c,\rm{onset}}\sim$ 48 K (ref. \mycite{ChenXH1}).
If assuming all carriers come from the electron pockets near the $M$ point, which is indeed the case in all other HED FeSe-based SCs and inferred by the Hall measurement in LiFeSe-122, the results summarized in Fig. 5(b) indicate that the high $T_{c}$ superconductivity emerges and is robust against the size variety of electron pockets once the electron doping level is larger than 0.05 e/Fe. One exception is A$_{x}$Fe$_{2-y}$Se$_{2}$, which shows similar electron doping level to LiFeSe-122 but with much lower $T_{c}$ ($\sim$ 30 K) \cite{Zhao2}. Such low $T_{c}$ could be related to the existence of Fe vacancies in FeSe layers, possibly detrimental to superconductivity. It has to be mentioned that once the FeSe layers in A$_{x}$Fe$_{2-y}$Se$_{2}$ are fully filled, another SC phase coexisting with the 30 K SC phase will appear and the $T_{c}$ will drastically increase to 44 K \cite{ZhangAM}. As for m-FeSe/STO films, it has the highest $T_{c}$ and the carrier doping level is in those of LiFeSe-122 and FeSe-11111. This implies that either there exists an optimal doping level or other factors may have influences on superconductivity. For example,
%the in-plane lattice parameter of m-FeSe/STO films is much larger than those in LiFeSe-122 and FeSe-11111 (refs. \mycite{Dong} and \mycite{TanSY}) and it means that the local environment of FeSe tetrahedron (bond lengths and/or bond angles) may be different in those compounds. Moreover,
the interfacial coupling between FeSe film and STO substrate should still be important to boost the $T_{c}$ to unprecedented value \cite{WangQY,LeeJ,Zheng}.

\section{Conclusion}

In summary, we have successfully synthesized the HED FeSe-based SC LiFeSe-122 single crystals. Our current observations in LiFeSe-122 single crystal without materials complexity suggests that the superconductivity in HED FeSe-based SCs is mainly determined by electron doping level and the suppression of hole pockets would closely related to the giant enhancement of $T_{c}$ to 40 - 50 K. Moreover, once above the threshold value (0.05 e/Fe), the $T_{c}$ seems insensitive to the carrier concentration and strong anisotropy of $\tau$, i.e., insensitive to the size and shape of electron pockets. These results will impose some constrains on theory study.
%In this compound, the intercalated Li atoms donate electrons to the FeSe layers and greatly increase the inter-FeSe-layer distance, leading to a superconductivity of 44 K and remarkable anisotropy of resistivity and $\mu_{0}H_{c2}$. Moreover, LiFeSe-122 shows significant temperature dependence of $n_{H}$ and cot$\theta_{H}$ at high temperature, possibly ascribed to the strong anisotropy of $\tau$ accompanying with possible variation of carrier concentration with temperature.
Because many kinds of AM as well as organic molecules can be intercalated in between FeSe layers and the electron doping level, crystallographic structure and the strength of inter-FeSe-layer coupling can be fine tuned by choosing suitable combinations of AM and organic molecules, this superconducting family provides an unique and novel platform for studying the origin of high $T_{c}$ superconductivity in FeSe-based SCs.

\section*{Acknowledgements}

This work was supported by the Ministry of Science and Technology of China (2016YFA0300504), the National Natural Science Foundation of China (No. 11574394), the Fundamental Research Funds for the Central Universities, and the Research Funds of Renmin University of China (RUC) (15XNLF06, 15XNLQ07). R.Y. was supported by the Ministry of Science and Technology of China (Grant No. 2016YFA0300504), the National Natural Science Foundation of China (Grant No. 11374361, 11674392).

%\section{Author contributions}

% H.C.L. provided strategy and designed the research. S.S.S. and S.H.W. performed the sample fabrication, measurements and fundamental data analysis; S.S.S., R.Y., and H.C.L. wrote the manuscript based on discussions with all the authors.

%\section{Additional information}

%\textbf{Supplementary Information} accompanies this paper.

%\textbf{Competing financial interests}: The authors declare that they have no competing financial interests.

$\dag$ These authors contributed equally to this work.

$\ast$ Correspondence and requests for materials should be addressed to H. C. Lei (email: hlei@ruc.edu.cn).

\end{document}